\documentclass[article,preprint,groupedaddress,12pt]{revtex4}
\usepackage{epsfig}
\usepackage{graphicx}
\usepackage{amssymb}

\begin{document}
\title{Strong Cosmic Censorship and the Universal Relaxation Bound}
\author{Shahar Hod}
\affiliation{The Ruppin Academic Center, Emeq Hefer 40250, Israel}
\affiliation{ }
\affiliation{The Hadassah Institute, Jerusalem 91010, Israel}
\date{\today}

\begin{abstract}
\ \ \ The strong cosmic censorship conjecture, introduced by Penrose
five decades ago, asserts that, in self-consistent theories of
gravity, Cauchy horizons inside dynamically formed black holes
should be unstable to remnant perturbation fields that fall into the
newly born black holes. The question of the (in)validity of this
intriguing conjecture in non-asymptotically flat charged black-hole
spacetimes has recently attracted much attention from physicists and
mathematicians. We here provide a general proof, which is based on
Bekenstein's generalized second law of thermodynamics, for the
validity of this fundamental conjecture in non-asymptotically flat
de Sitter black-hole spacetimes.
\newline
\newline
\end{abstract}
\bigskip
\maketitle

The influential singularity theorems of Hawking and Penrose
\cite{HawPen,Pen1,Pen2} have dramatically changed our understanding
of gravity and curved black-hole spacetimes. In particular, these
mathematically elegant theorems have revealed the physically
important fact that spacetime singularities, regions in which
classical general relativity loses its predictive power, may
naturally be formed by self-gravitating matter configurations whose
dynamical collapse is governed by the Einstein field equations.

Following this intriguing (and highly disturbing) observation,
Penrose \cite{Pen1,Pen2} has suggested that the deterministic nature
of classical theories of gravity is nevertheless preserved by a
mysterious ``cosmic censor'' which prevents spacetime singularities
from being observed by distant observers. In particular, the strong
version of the Penrose cosmic censorship conjecture asserts that
Cauchy horizons inside black holes, which mark the deterministic
boundaries of spacetime for dynamical evolutions of matter fields
which are governed by the classical Einstein field equations, are
singular and cannot be crossed by physical observers
\cite{Pen1,Pen2}.

It should be emphasized that, at present, there is no general
mathematically rigorous proof for the validity of the Penrose strong
cosmic censorship conjecture (SCCC) in curved spacetimes. In
particular, the evidence in favor (or against) the existence of a
mysterious cosmic censor who protects the deterministic power of
classical general relativity is based on our (rather limited)
experience in solving the non-linearly coupled Einstein-matter field
equations \cite{Cham,Hod1,Hod1c,Ge,Refin}.

In order to disprove the Penrose SCCC, many researches have tried
during the last five decades to identify curved black-hole
spacetimes which possess inner stable Cauchy horizons which are
regular enough to allow a continuation of the spacetime metric
beyond that horizons in a non-unique (that is, in a
non-deterministic) way. If exist, such black-hole spacetimes would
imply the catastrophic failure of the SCCC in some curved spacetimes
and the corresponding breakdown of determinism in classical general
relativity.

The question of the (in)validity of the Penrose SCCC in
non-asymptotically flat de Sitter black-hole spacetimes has
attracted much attention from physicists and mathematicians during
the last couple of years \cite{Refin}. In particular, it has
recently been shown \cite{Refin} that, in de Sitter black-hole
spacetimes, the fate (singular/non-singular) of the potentially pathological inner Cauchy
horizons is determined by a delicate competition between two
opposite physical mechanisms:
\newline
(1) The exponential late-time {\it decay}
$\psi^{\text{external}}(t\to\infty)\sim e^{-\Im\omega_0\cdot t}$ of
remnant matter fields which propagate in the external spacetime
regions of the dynamically formed black holes. The characteristic
rate of this decay mechanism is determined by the fundamental (least
damped) quasinormal resonant frequency $\omega_0$ of the composed
black-hole-field system.
\newline
(2) The exponential {\it amplification} (blue-shift) mechanism
$\psi^{\text{internal}}(v\to\infty)\sim e^{\kappa_- v}$
\cite{Notevv} which characterizes the dynamics of the remnant fields
as they fall into the newly born black holes and accumulate along
their inner (Cauchy) horizons. The characteristic rate of this
amplification mechanism is determined by the surface gravity
$\kappa_-$ of the inner black-hole horizon.

Specifically, the (in)stability properties of the inner Cauchy
horizons inside dynamically formed black holes in non-asymptotically
flat de Sitter spacetimes are determined by the dimensionless ratio
\cite{Refin}
\begin{equation}\label{Eq1}
\Gamma\equiv {{\Im\omega_0}\over{\kappa_-}}\  .
\end{equation}
In particular, dynamically formed black holes in asymptotically de
Sitter spacetimes which are characterized by the dimensionless
inequality $\Gamma>1/2$ are pathological from the point of view of
the Penrose SCCC in the sense that they contain inner Cauchy
horizons which are stable and regular enough to allow a continuation
of the spacetime metric in a {\it non}-unique (non-deterministic)
way \cite{Refin}.

Thus, a necessary condition for classical general relativity to preserve its predictive power in
non-asymptotically flat de Sitter black-hole spacetimes is given by the simple inequality
\begin{equation}\label{Eq2}
\Gamma\leq{1\over2}\  .
\end{equation}
For the advocates of the Penrose SCCC, the task remains to prove that
the resonant relaxation spectra of {\it all} dynamically formed
black holes in non-asymptotically flat curved spacetimes possess (at
least) one quasinormal resonant frequency with the property [see
Eqs. (\ref{Eq1}) and (\ref{Eq2})]
\begin{equation}\label{Eq3}
\Im\omega_0\leq{1\over2}\kappa_-\  .
\end{equation}

Interestingly, using analytical techniques, it has been explicitly
proved in \cite{Hod2} that Kerr black holes in asymptotically de
Sitter spacetimes conform to the inequality (\ref{Eq3}). These
spinning curved spacetime solutions of the Einstein field equations therefore respect determinism.

As for the case of charged black holes in asymptotically de Sitter
spacetimes, the situation is more involved \cite{Cham,Hod1,Hod1c,Ge,Refin}. In particular, it has
been proved \cite{Hod1} that the resonant relaxation spectra of composed
charged-Reissner-Nordstr\"om-de-Sitter-black-holes-charged-massive-fields
systems in the dimensionless physical regime $\mu r_+ \ll qQ \ll
(\mu r_+)^2$ [here $\{Q,r_+\}$ are respectively the electric charge
and the outer horizon radius of the black hole and $\{\mu,q\}$ are
respectively the proper mass and charge coupling constant of the
matter field] are characterized by the fundamental inequality (\ref{Eq3}). These charged
black-hole-field systems therefore respect the Penrose SCCC.

It is clear, however, that the proof presented in \cite{Hod1} for
the validity of the inequality (\ref{Eq3}) \cite{Notenec} in the
regime $\mu r_+ \ll qQ \ll (\mu r_+)^2$ is not enough. In
particular, one would like to have a generic (that is,
parameter-{\it independent}) proof that the resonant relaxation
spectra of {\it all} dynamically formed Reissner-Nordstr\"om-de
Sitter (RNdS) black holes are characterized by the fundamental
inequality (\ref{Eq3}) which acts as a necessary condition for the
validity of the SCCC in these non-asymptotically flat curved
spacetimes.

Due to the complexity of the non-linearly coupled
Einstein-charged-matter field equations, any attempt to obtain a
general proof for the validity of the Penrose SCCC in charged curved
spacetimes seems, at least at first glance, hopeless. In particular,
it should be realized that a direct brute force test of the
(in)validity of the Penrose SCCC in charged de Sitter spacetimes
would require one to study (probably numerically) the non-linear
dynamics of self-gravitating charged matter fields that collapse to
form charged black holes. Specifically, one should scan the entire
(infinitely large) phase space of the physical parameters
$\{r_-,r_+,r_{\text{c}},q,\mu\}$ \cite{Noteradii} which characterize
the newly born black-hole spacetimes in order to determine whether
or not there exist dynamically formed RNdS black holes that violate
the fundamental inequality (\ref{Eq3}) \cite{Notenec,Notephys}. This
direct approach to the problem seems to be a truly Sisyphean task!

But we need not lose heart -- one should always remember that the
basic physical laws of nature can provide, in a relatively
straightforward way, general insights about the physical properties
of relatively complex systems. One such law is Bekenstein's
generalized second law of thermodynamics \cite{Bek1} which asserts
that the total entropy of a black-hole spacetime (that is, the
entropy $S_{\text{BH}}=A/4\hbar$ \cite{NoteA} of the black hole
itself plus the entropy of the external matter fields) is an
irreducible quantity in any physically self-consistent quantum
theory of gravity.

Interestingly, and most importantly for the question of the
(in)validity of the Penrose SCCC in black-hole spacetimes, it has
been explicitly shown in \cite{Hod3} that Bekenstein's generalized
second law of thermodynamics yields, in a rather
straightforward way, a fundamental upper bound on the characteristic
relaxation rates of dynamical physical systems with a well defined
thermodynamic description. Specifically, for a thermodynamic system
which is characterized by a temperature $T$, the universal
relaxation bound can be expressed by the remarkably compact
time-times-temperature (TTT) quantum relation \cite{Hod3}
\begin{equation}\label{Eq4}
\tau\times T\geq {{\hbar}\over{\pi}}\  ,
\end{equation}
where $\tau$ is the characteristic relaxation time of the
thermodynamic system.

Remarkably, curved black-hole spacetimes are known to be
characterized by a well defined temperature, the Bekenstein-Hawking
temperature \cite{Bek1,Haw1}
\begin{equation}\label{Eq5}
T_{\text{BH}}={{\kappa_+}\over{2\pi}}\cdot\hbar\  .
\end{equation}
Taking cognizance of the universal relaxation bound (\ref{Eq4}) and
the quantum Bekenstein-Hawking relation (\ref{Eq5}) one immediately
deduces that the resonant relaxation spectra of {\it all}
dynamically formed black holes are characterized by the simple upper
bound \cite{Notetw}
\begin{equation}\label{Eq6}
\Im\omega_0\leq {1\over2} \kappa_+\  ,
\end{equation}
where $\omega_0$ is the fundamental (least damped) resonant
frequency of the composed
dynamically-formed-black-hole-remnant-field system.

What we find most interesting is the fact that, in black-hole
physics where $T_{\text{BH}}\propto\hbar$ [see Eq. (\ref{Eq5})],
the {\it quantum} lower bound (\ref{Eq4}) becomes a {\it classical}
(that is, $\hbar$-independent) relation between the characteristic
relaxation rates of the dynamically formed black-hole spacetimes and
the corresponding black-hole surface gravities.

%\section{Summary}
{\it Summary.---} The conjectured strong cosmic censorship principle
\cite{HawPen,Pen1,Pen2} has attracted much attention from physicists
and mathematicians since its introduction by Penrose five decades
ago. In particular, the question of the (in)validity of this highly
influential conjecture in charged de Sitter black-hole spacetimes
has been the focus of an intense debate in the last couple of years
(see \cite{Cham,Hod1,Hod1c,Ge} and references therein).

The SCCC asserts that, in self-consistent theories of gravity, spacetime metrics inside physically
realistic ({\it dynamically} formed) black holes cannot be extended
in a non-unique (non-deterministic) way beyond the inner Cauchy
horizons. For this to happen, the inner Cauchy horizons of
dynamically formed black-hole spacetimes should be unstable to
remnant perturbation fields that fall into the newly born black
holes. In particular, a necessary condition for the validity of the
SCCC in asymptotically de Sitter black-hole spacetimes is provided
by the simple inequality (\ref{Eq3}).

A direct way to prove the validity of the fundamental inequality
(\ref{Eq3}) would be to scan numerically the infinitely large phase
space of the resonant spectra which characterize the relaxation of
dynamically formed black-hole spacetimes. Instead of following this
truly Sisyphean (and numerically time consuming) task, we have
pointed out that the Bekenstein generalized second law of
thermodynamics \cite{Bek1}, a physical law which is widely believed
to reflect a fundamental aspect of a physically self-consistent
quantum theory of gravity, yields the upper relaxation bound
(\ref{Eq6}) \cite{Hod3,Notephys2,Schsh}. Using the characteristic
inequality $\kappa_+\leq \kappa_-$ for the surface gravities of the
black-hole spacetime \cite{Cham}, one immediately realizes that the
relaxation spectra of all dynamically formed black holes are indeed
characterized by the desired inequality (\ref{Eq3}).

We therefore conclude that physically realistic
\cite{Notephys2,Schsh} black holes in asymptotically de Sitter
spacetimes {\it respect} determinism and conform to the Penrose
strong cosmic censorship conjecture.

\bigskip
\noindent
{\bf ACKNOWLEDGMENTS}
\bigskip

This research is supported by the Carmel Science Foundation. I would
like to thank Yael Oren, Arbel M. Ongo, Ayelet B. Lata, and Alona B.
Tea for helpful discussions.


\begin{thebibliography}{99}
\bibitem{Notevv} We use gravitational units in which $G=c=1$. Here $v$ is the standard advanced null coordinate.
\bibitem{Notenec} It is worth emphasizing again that the inequality (\ref{Eq3}) serves
as a necessary condition for the validity of the Penrose SCCC in de
Sitter black-hole spacetimes.
\bibitem{Noteradii} Here $\{r_-,r_+,r_{\text{c}}\}$ are respectively
the inner (Cauchy), outer (event), and cosmological horizons of the
non-asymptotically flat RNdS black-hole spacetime.
\bibitem{NoteA} Here $A$ is the black-hole surface area.
\bibitem{Notetw} Here we have used the relation $\tau_{\text{relax}}\equiv
1/\Im\omega_0$ for the characteristic relaxation time of the dynamically formed black-hole spacetime,
where $\omega_0$ is the fundamental (least damped) resonant frequency of the composed
black-hole-remnant-field system.
\end{thebibliography}
\end{document}